\documentclass[showkeys,showpacs]{revtex4}
\usepackage{graphicx}
\usepackage{amsmath}
\usepackage{amssymb}
\usepackage{color}

\begin{document}
\title{Colored dark matter}

\author{Vladimir Dzhunushaliev
\footnote{Senior Associate of the Abdus Salam ICTP}}
\email{dzhun@krsu.edu.kg}
\affiliation{Dept. Phys. and Microel.
Engineer., Kyrgyz-Russian Slavic University, Bishkek, Kievskaya Str.
44, 720021, Kyrgyz Republic}

\begin{abstract}
The idea is considered that a classical non-Abelian gauge field can be considered as a dark matter candidate. It is shown that Yang-Mills equations have solutions with such distribution of the mass density that allows to describe a rotational curve of spiral galaxies. The conditions necessary for such consideration are considered. One parameter is estimated from Yang-Mills equations and from astrophysical observations (stars rotation curve). The agreement is to within 1\%.
\end{abstract}

\date{\today}

\maketitle

\section{Introduction}

The electromagnetic field behaves at different conditions either as pure quantum or almost as a classical field. But for non-Abelian gauge fields we know only quantum manifestations of these fields. Here we would like to present the idea that classical non-Abelian gauge fields can become visible as DM.
\par 
The problem of the DM nature is now one of the most  fundamental problems in modern physics. The reason is to explain the invisibility of DM it is necessary to involve such kinds of an exotic matter as, for example, WIMP, supersymmetric particles etc. 
\par 
In this article we offer a new model of DM. In the studied model the DM is an interior of an abortive singularity or black hole. We understand this suggestion as follows. 
\par
In Ref's~\cite{Obukhov:1996ry}~\cite{Dzhunushaliev:1999fy} it is shown that the energy density of spherically symmetric solutions of Yang-Mills equations may have non-standard behavior (behind an exception, of course,  't Hooft - Polyakov monopole). For example, in Ref.~\cite{Dzhunushaliev:1999fy} it is shown that the energy density is weakly decreasing. It is natural that such distribution of the matter will give diverging solutions by switching on the gravity. For some boundary conditions at the center it will be an interior of a non-Abelian back hole and for other values of the boundary conditions a singularity located at some distance from the center will appear. But one very interesting subtlety exists here. This solutions are classical ones. The careful analysis shows that at the infinity these solutions are strongly  oscillating  in the space (gravity is switched off) and the period of the oscillations increases by moving away from the centre. It is obvious that on some distance from the center the quantum fluctuations of gauge field on the period distance of oscillations become comparable with a field magnitude. It means that at this distance the gauge field becomes quantum one. The problem in calculating of such  distribution of the non-Abelian gauge field is that the gauge field becomes essentially non-perturbative and it is impossible to apply the perturbative Feynman diagram techniques to its description. 
\par 
The idea presented here is that the radius where the gauge field becomes quantum can be less than the radius of a singular or an event horizon corresponding to a non-Abelian black hole. In this case this object does become neither a singularity nor a black hole (abortive singularity/black hole). It looks as follows: in space  there is a sphere filled with a classical gauge field and on border of this sphere the non-Abelian gauge field becomes quantum one and the mass contained in this sphere is not enough for formation of a singularity or an event horizon. The color gauge field in the sphere does not interact with the elementary particles because the particles are colorless and consequently this classical field is invisible and can be applied as a candidate of DM. 
\par 
In fact in this paper we consider the idea that in a non-Abelian gauge theory may exist such space distribution of a gauge field that classical and quantum phases exist simultaneously but spatially separated. In this case the jump condition from classical phase to quantum one is the strong oscillations of classical non-Abelian field. These oscillations leads to the fact that quantum fluctuations become essential at a distance of oscillation period. In other words the space is filled with a non-perturbative gauge vacuum where there exist defects filled with the  classical gauge field. The galaxies are located in these defects and the classical gauge field is the DM. 

\section{The interior of an Einstein-Yang-Mills singularity/black hole}

In this section we would like to show that usually the SU(3) gauge field distribution leads to a singular spacetime. We use the following metric
\begin{equation}
	ds^2 = e^{\nu(r)} \left[ 1 - \frac{M(r)}{r} \right] dt^2 - 
	\frac{dr^2}{1 - \frac{M(r)}{r}} - 
	r^2 \left(
		d \theta^2 + \sin^2 \theta d \varphi^2
	\right)
\label{1-10}
\end{equation}
where $t,r,\theta,\varphi$ are usual spherical coordinates. Substituting metric \eqref{1-10} and SU(3) gauge potential \eqref{a1-10}-\eqref{a1-60} into Einstein-Yang-Mills equations 
\begin{eqnarray}
	R_{\mu \nu} - \frac{1}{2} g_{\mu \nu} R &=& \varkappa T_{\mu \nu},
\label{1-50}\\
	D_\nu F^{a \mu \nu} &=& 0, 
\label{1-55}\\
	T_{\mu \nu} &=& -F^a_{\mu \alpha} F_\nu^{\;a \alpha} + 
		\frac{1}{4} g_{\mu \nu} F^a_{\alpha \beta}F^{a \alpha \beta}
\label{1-60}
\end{eqnarray}
gives us Einstein equations 
\begin{eqnarray}
	M' &=& \frac{\varkappa}{r^2} \left\{
		\frac{2}{3} \frac{e^{-\nu}}{1 - \frac{M}{r}} \left[
			6v^2 w^2 + \left( 1 - \frac{M}{r} \right) \left( w - r w' \right)^2 
		\right] + 
		2 \left[
			\left( 1 - v^2 \right)^2 + 2 r^2 \left( 1 - \frac{M}{r} \right) {v'}^2
		\right]
	\right\}	,
\label{1-70}\\
	r \left( 1 - \frac{M}{r} \right) \nu' &=& \frac{8 \varkappa}{r^2} \left[
		\frac{e^{-\nu}}{1 - \frac{M}{r}} v^2 w^2 + r^2 \left( 1 - \frac{M}{r} \right) {v'}^2
	\right],
\label{1-80}\\
	- \frac{\nu ''}{2} \left( 1 - \frac{M}{r} \right) &+& \frac{M''}{2r} - 
	\frac{{\nu'}^2}{4} \left( 1 - \frac{M}{r} \right) + \frac{3}{4} \frac{M' \nu'}{r} - 
	\frac{\nu'}{4r} \left( 2 + \frac{M}{r} \right)
	= 
	- \frac{2}{3} \frac{\varkappa}{r^4} \left[
		3 \left( 1 - v^2 \right)^2 + e^{-\nu} \left( w - r w' \right)^2
	\right]
\label{1-90}
\end{eqnarray} 
and Yang-Mills equations 
\begin{eqnarray}
	\left( 1 - \frac{M}{r} \right) v'' + 
	\left(
		- \frac{M'}{r} + \frac{M}{r^2}
	\right) v' &=& 
	\frac{v}{r^2} \left( v^2 - 1 \right) - 
	\frac{e^{-\nu}}{r^2 \left( 1 - \frac{M}{r} \right)} v w^2 , 
\label{1-94}\\
	w'' - w' \nu ' + w \frac{\nu'}{r} &=& \frac{6}{r^2 \left( 1 - \frac{M}{r} \right)} 
	w v^2
\label{1-98}
\end{eqnarray} 
where $\varkappa$ is the gravitational constant, $a=1,2,\cdots , 8$ is the color index and $\chi(r) = h(r) = 0$. The analytical solution does not exist and we search the numerical solutions for this equations set. 
\par
\begin{figure}[h]
\begin{minipage}[t]{.45\linewidth}
  \begin{center}
  \fbox{
  \includegraphics[height=5cm,width=7cm]{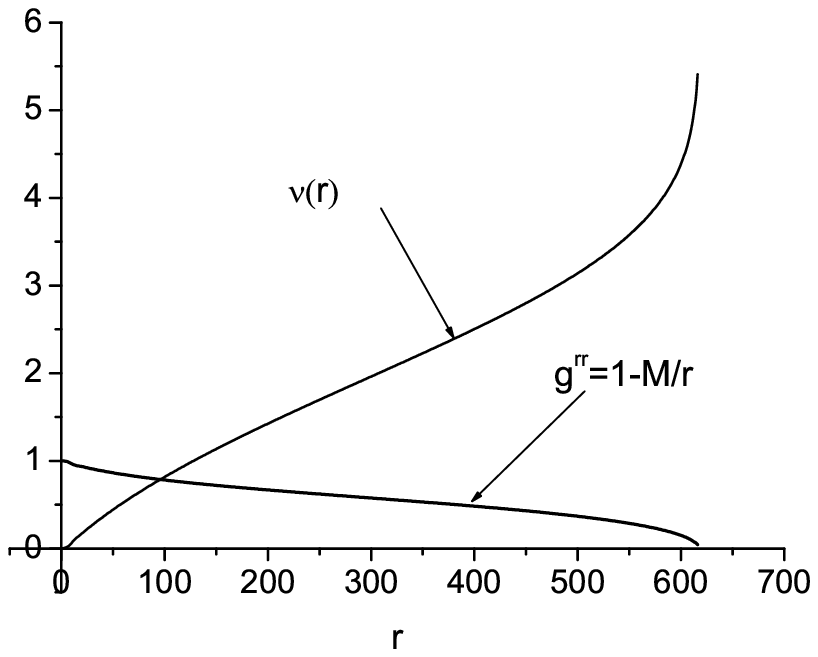}}
  \caption{The profile of the functions $g^{rr} = 1 - \frac{M(r)}{r}$ and $\nu(r)$.}
  \label{event_horizon}
  \end{center}
\end{minipage}\hfill
\begin{minipage}[t]{.45\linewidth}
  \begin{center}
  \fbox{
  \includegraphics[height=5cm,width=7cm]{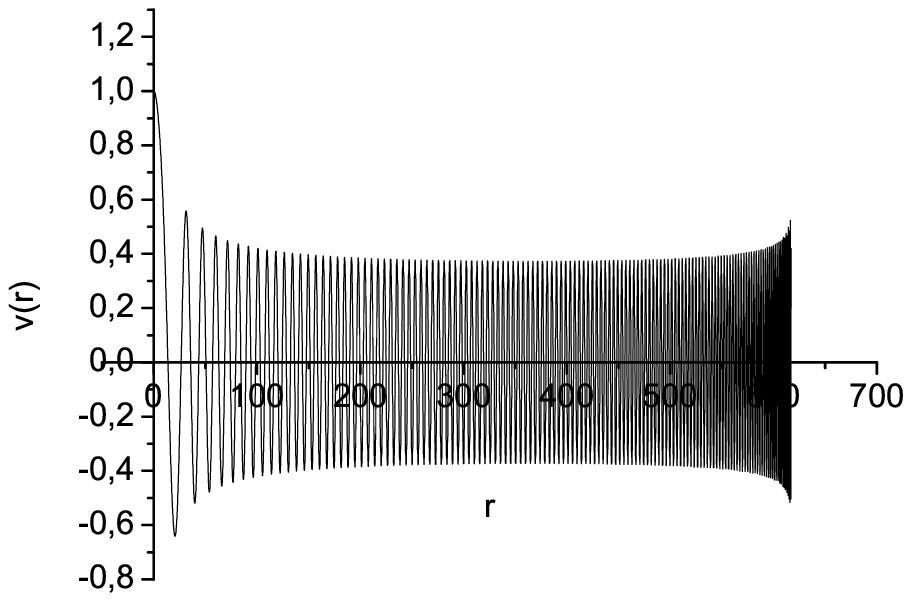}}
	  \caption{The profile of the function $v(r)$.}
  \label{v_r}
  \end{center}
\end{minipage}\hfill 
\end{figure}

\begin{figure}[h]
\begin{minipage}[t]{.45\linewidth}
  \begin{center}
  \fbox{
  \includegraphics[height=5cm,width=7cm]{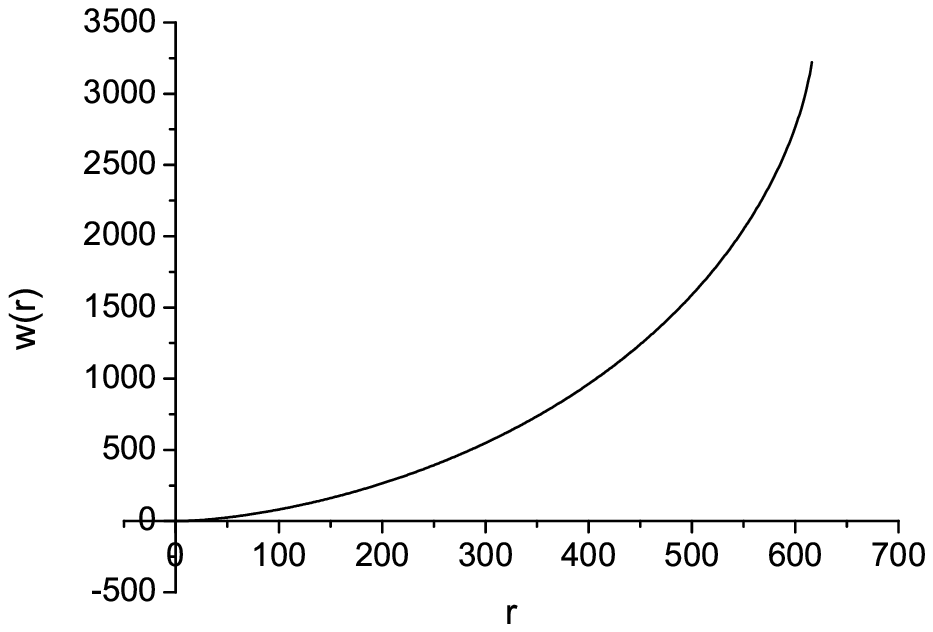}}
  \caption{The profile of the function $w(r)$.}
  \label{w_r}
  \end{center}
\end{minipage}\hfill
\begin{minipage}[t]{.45\linewidth}
  \begin{center}
  \fbox{
  \includegraphics[height=5cm,width=7cm]{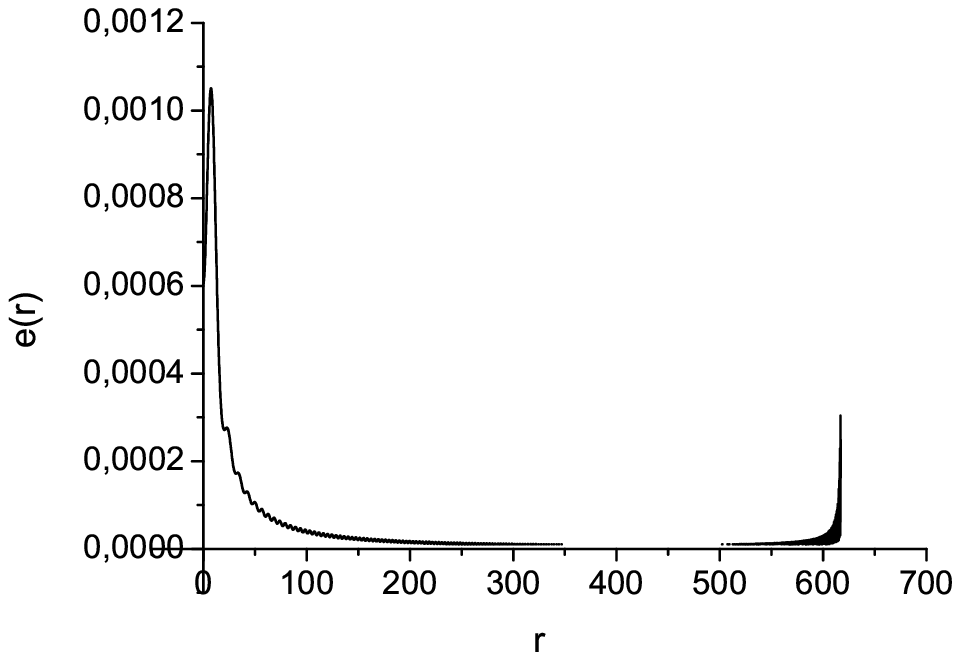}}
	  \caption{The profile of the function energy density.}
  \label{energy_density}
  \end{center}
\end{minipage}\hfill 
\end{figure}

 
We search for the solution inside of the non-Abelian singularity/black hole. It means that we should start the solution from the point $r=0$. The analytical solution close to the center is 
\begin{eqnarray}
	M(r) &=& M_3 \frac{r^3}{6} + \mathcal O \left( r^5 \right), \quad 
	\nu(r) = \nu_2 \frac{r^2}{2} + \mathcal O \left( r^4 \right), 
\label{1-100}\\
	v(r) &=& 1 + v_2 \frac{r^2}{2} + \mathcal O \left( r^4 \right), \quad 
	w(r) = w_3 \frac{r^3}{6} + \mathcal O \left( r^5 \right), 
\label{1-110}\\
	M_3 &=& 5 \varkappa v_2^2, \quad 
	\nu_2 = 8 \varkappa v_2.  
\label{1-120}
\end{eqnarray}
The numerical solution can not start from the point $r=0$ since Eq's \eqref{1-70}-\eqref{1-90} have terms $1/r^2$. Consequently we should start from the point $x = \delta \ll r/\sqrt \varkappa$. The boundary conditions are 
\begin{eqnarray}
	M(\delta) &=& M_3 \frac{\delta^3}{6} , \quad 
	\nu(\delta) = \nu_2 \frac{\delta^2}{2} , 
\label{1-130}\\
	v(\delta) &=& 1 + v_2 \frac{\delta^2}{2} , \quad 
	v'(\delta) = v_2 \delta , 
\label{1-140}\\
	w(\delta) &=& w_3 \frac{\delta^3}{6}, \quad 
	w'(\delta) = w_3 \frac{\delta^2}{2}.
\label{1-150}
\end{eqnarray}
We present the profiles of the functions $g^{rr} = ( 1 - \frac{M(r)}{r} ), \nu(r), v(r), w(r)$ in Fig's~\ref{event_horizon} - \ref{w_r}. 
We see that at a point $r = r_H$ the function 
$g^{rr}\left( r_H \right) = 1 - \frac{M(r_H)}{r_H} = 0$. The energy density is 
\begin{equation}
\begin{split}
	\varepsilon = T^0_0 = & -F_{0 \alpha} F^{0 \alpha} + 
	\frac{1}{4} F_{\alpha \beta}F^{\alpha \beta} 
	= \frac{2}{3g^2} 
		\frac{e^{-\nu}}{ r^4 \left( 1 - \frac{M}{r} \right)} \left[
			6v^2 w^2 + \left( 1 - \frac{M}{r} \right) \left( w - r w' \right)^2 
		\right] + 
\\
	& 
		\frac{2}{r^4} \left[
		\left( 1 - v^2 \right)^2 + 2 r^2 {v'}^2
		\right]
\end{split}
\label{1-160}
\end{equation}
and its profile in Fig.~\ref{energy_density} is presented. 
Consequently the numerical analysis have shown us that at the point $r_H$ where $1 - M(r_H)/r_H = 0$ there is a real or coordinate singularity (for some boundary conditions it will be an event horizon). From Fig.~\ref{v_r} we see that the function $v(r)$ is oscillating function and the period of oscillations increases with the moving away from the center. At some distance $r_q$ from the center the quantum fluctuations on the distance of period will be comparable with the field magnitude. In this case the classical gauge field becomes quantum one. If it occurs on the distance $r_q < r_H$ then we will have a ball filled with the classical gauge field inside quantum gauge field. In Ref.~\cite{Dzhunushaliev:2006di} is claimed that the non-perturbative quantized SU(3) gauge field can be described using two scalar fields. These fields decrease very quickly (exponentially) to a ground state. If this so then the situation looks as follows: in the space there is a ball filled with the classical gauge field with weakly decreasing energy density. At some distance $r_q$ from the center the field becomes quantum one and the energy density very quickly (exponentially) decreases to a ground state (non-perturbative vacuum). The idea presented here is that the ball can be considered as the DM as usual elementary particles are colorless and consequently can not interact with the color classical SU(3) gauge fields. 
\par 
Let us note that non-Abelian black holes and particlelike solutions exist only with a special choice of the boundary conditions on the event horizon or at the center. We have to underline that in contrast with the non-Abelian black holes and particlelike solutions we consider the solutions with arbitrary boundary conditions. 
\par 
In the following sections we will estimate the radius $r_H$ of a real/coordinate singularity and the radius $r_q$ where the transition from the classical phase to quantum one occurs. 

\section{The estimation of the radius of real or coordinate singularity}

Unfortunately the solution presented in previous section is numerical one that does not allow us to calculate the sphere where the above mentioned singularity (real or coordinate) is. We will estimate the radius of a singular sphere (using Newton gravity) in the following way. The radius is estimated as the place where the Newtonian gravitational potential becomes so strong that a test particle should have a velocity of light to get from this point on infinity. We test this method for the Schwazschild black hole solution. The energy conservation law tells us 
\begin{equation}
	\frac{m c^2}{2} - G \frac{m M}{r_H} = 0
\label{2-10}
\end{equation}
here $G$ is the Newtonian constant; the first and second terms are kinetic and potential energies of a test particle with the mass $m$ and $M$ is the mass of a singularity or a black hole. As a result we have the radius of a singularity or an event horizon 
\begin{equation}
	r_H = \frac{2 G M}{c^2} 
\label{2-20}
\end{equation}
that absolutely precisely coincides with the event horizon radius calculated in general relativity. 
\par 
Now we will try to estimate the radius of the singularity or the event horizon for the classical distribution of the SU(3) gauge fields \eqref{a1-80}-\eqref{a1-110}. We will work in Newton gravity because the values $v_2$ and $w_3$ are small enough, i.e. the magnitudes of the SU(3) gauge fields close to the center are small enough. In this case Yang-Mills equations in a flat space are 
\begin{equation}
	D_\nu F^{a\mu \nu} = 0 
\label{2-30}
\end{equation}
and with ansatz \eqref{a1-80}-\eqref{a1-110} where $\chi(r) = h(r) = 0$ we have the following equations 
\begin{eqnarray}
	x^2 w'' &=& 6w v^2 ,
\label{2-40}\\
	x^2 v'' &=& v^3 - v - v w^2 
\label{2-50}
\end{eqnarray}
here the dimensionless radius $x = r/r_0$ is introduced and $r_0$ is an arbitrary constant. In fact Eq's \eqref{2-40} \eqref{2-50} are Eq's \eqref{1-94} \eqref{1-98} in Minkowski spacetime. The asymptotical behavior $x \gg 1$ of the solution is \cite{Dzhunushaliev:1999fy} 
\begin{eqnarray}
	v(x) &\approx& A \sin \left( x^\alpha + \phi_0 \right),
\label{2-60}\\
	w(x) &\approx& \pm \left[
		\alpha x^\alpha + \frac{\alpha - 1}{4} 
		\frac{\cos \left( 2 x^\alpha  + 2 \phi_0 \right)}{x^\alpha}
	\right] ,
\label{2-70}\\
	3 A^2 &=& \alpha (\alpha - 1)
\label{2-80}
\end{eqnarray}
with $\alpha > 1$. The energy density $\epsilon(x)$ is 
\begin{equation}
	\epsilon(r) = 
	- F^a_{0i} F^{a0i} + \frac{1}{4} F^a_{ij} F^{aij} = 
	\frac{1}{g^2 r_0^4} \left[ 
	4 \frac{{v'}^2}{x^2} + 
	\frac{2}{3} \frac{\left( x w' - w \right)^2}{x^4} + 
	2 \frac{\left( v^2 - 1 \right)^2}{x^4} + 
	4 \frac{ v^2 w^2}{x^4} 
	\right] = 
	\frac{1}{g^2 r_0^4} \varepsilon(x)
\label{2-90}
\end{equation}
where $v'=dv/dx, w'=dw/dx$. Asymptotically the dimensionless energy density is 
\begin{equation}
	\varepsilon_\infty(x) \approx 
	\frac{2}{3} 
	\alpha^2 \left( \alpha - 1 \right) \left( 3 \alpha - 1 \right) 
	\left( \frac{r}{r_0} \right)^{2 \alpha - 4}. 
\label{2-100}
\end{equation}
The numerical analysis shows that the functions $v(r)$ and $w(r)$ quickly attain an asymptotic form. Therefore we will use \eqref{2-100} to estimate the mass $\mathcal M(r)$ under radius $r$
\begin{equation}
	\mathcal M(r) = \frac{4 \pi}{c^2} \int\limits_0^r r^2 \varepsilon(r) dr \approx 
	\frac{8 \pi}{3 g^2 r_0 c^2} \frac{\alpha^2 (\alpha-1) (3 \alpha-1)}{2 \alpha - 1} 
	\left( \frac{r}{r_0} \right)^{2\alpha -1}.
\label{2-110}
\end{equation}
The same calculations as in \eqref{2-10} gives us 
\begin{equation}
	\frac{m c^2}{2} - G \frac{m \mathcal M(r_H)}{r_H} = 0
\label{2-120}
\end{equation}
that leads to  
\begin{equation}
	r_H \approx r_0 \left[
		\frac{3}{4} {g'}^2 \frac{2 \alpha - 1}{\alpha^2 (\alpha - 1) (3 \alpha - 1)} 
		\left( \frac{r_0}{l_{Pl}} \right)^2 
	\right]^{\frac{1}{2\alpha - 2}}
\label{2-130}
\end{equation}
where $\frac{1}{{g'}^2} = \frac{4 \pi /g^2}{\hbar c}$ is the dimensionless coupling constant in the SU(3) gauge theory; $1/g$ is the analog of the electric charge in electrodynamics; 
$l_{Pl} = \sqrt{\frac{\hbar G}{c^3}} \approx 10^{-35} $ m is the Planck length. The numerical factor 
\begin{equation}
	\frac{3}{4} {g'}^2 \frac{2 \alpha - 1}{\alpha^2 (\alpha - 1) (3 \alpha - 1)} 
	\approx 1
\label{2-140}
\end{equation}
and consequently 
\begin{equation}
	r_H \approx r_0 \left( \frac{r_0}{l_{Pl}} \right)^{\frac{1}{\alpha - 1}}. 
\label{2-150}
\end{equation}
Using the parameters $r_0$ and $\alpha$ which are calculated in Appendixes~\ref{alpha} and \ref{asymptotic} we have 
\begin{equation}
	r_H \approx 10^{28} \mathrm{m} 
\label{2-155}
\end{equation}
that is much more that the galaxy radius. 
\par
Now we can define the natural choice of the parameter $r_0$. At the origin Eq's \eqref{2-40} \eqref{2-50} have the solution that can be presented as series \eqref{1-110} 
\begin{eqnarray}
	v(r) &=& 1 + \frac{1}{2} \left( r_0^2 v_2 \right) \left( \frac{r}{r_0} \right)^2 + 
	\mathcal O \left[ \left( \frac{r}{r_0} \right)^4 \right] = 
	1 + v'_2 \frac{x^2}{2}  + \mathcal O \left( x^4 \right) , 
\label{2-160}\\
	w(r) &=&  \frac{1}{6} \left( r_0^3 w_3 \right) \left( \frac{r}{r_0} \right)^3 + 
	\mathcal O \left[ \left( \frac{r}{r_0} \right)^5 \right] = 
	w'_3 \frac{x^3}{6} + \mathcal O \left( x^5 \right).
\label{2-170}
\end{eqnarray}
Therefore the natural choice of the parameter $r_0$ is 
\begin{equation}
	\text{either} \quad r_0^2 = \frac{1}{v_2} \quad
	\text{or} \quad 
	r_0^3 = \frac{1}{w_3}.
\label{2-180}
\end{equation}

\section{The transition from classical phase to quantum one}
\label{transition}

The idea considered above mentioned solution of the classical Yang-Mills equations can not be extended up to infinity because in some place the space oscillations of the classical gauge field becomes so strong that non-perturbative quantum effects should be taken into account. In this section we try to estimate the radius where it can happen. 
\par
Following on the Heisenberg uncertainty principle 
\begin{equation}
	\frac{1}{c} \; \Delta F^a_{ti} \; \Delta A^{a i} \; \Delta V \approx \hbar
\label{3-10}
\end{equation}
here $\Delta F^a_{ti}$ is a quantum fluctuation of color electric field $F^a_{ti}$; $\Delta A^{a i}$ is a quantum fluctuation of color electric potential $A^{a i}$; $\Delta V$ is the volume where the quantum fluctuations $\Delta F^a_{ti}$ and $\Delta A^{a i}$ takes place; $a=1,2 \cdots 8$ is the color index; $i=1,2,3$ is the space index. 
\par 
For the ansatz \eqref{a1-80}-\eqref{a1-110}
\begin{equation}
	F^2_{t \theta} = - \frac{2}{g} \sin \theta \frac{v w}{r} .
\label{3-20}
\end{equation}
We introduce the physical component of the $F^2_{t \theta}$ 
\begin{equation}
	\left| \tilde F^2_{t \theta} \right| = 
	\sqrt{F^2_{t \theta} F^{2\theta}_t} = \frac{2}{g} \sin \theta \frac{v w}{r^2} .
\label{3-30}
\end{equation}
To an accuracy of a numerical factor the fluctuations of the SU(3) color electric field are 
\begin{equation}
	\Delta \tilde F^2_{t \theta} \approx \frac{1}{g} \frac{1}{r^2} \left(
		\Delta v \; w + v \; \Delta w
	\right).
\label{3-40}
\end{equation}
For the ansatz \eqref{a1-80}-\eqref{a1-110}
\begin{eqnarray}
	A^2_\theta = 0,
\label{3-70}\\
	A^{1,3,4,6,8}_\theta \approx \frac{1}{g} v.
\label{3-80}
\end{eqnarray}
Introducing the physical components of the gauge potential $A^{1,3,4,6,8}_\theta$ 
\begin{equation}
	\left| \tilde A^{1,3,4,6,8}_\theta \right| = 
	\sqrt{A^{1,3,4,6,8}_\theta A^{1,3,4,6,8; \theta}} \approx 
	\frac{1}{g} \sin \theta \frac{v}{r} 
\label{3-90}
\end{equation}
we assume that 
\begin{equation}
	\Delta \tilde A^2_\theta \approx \Delta \tilde A^1_\theta \approx 
	\frac{1}{g} \sin \theta \frac{\Delta v}{r} .
\label{3-100}
\end{equation}
The volume $\Delta V$ is 
\begin{equation}
	\Delta V = 4\pi r^2 \Delta r .
\label{3-110}
\end{equation}
The period of space oscillations by $r \gg r_0$ can be defined in the following way 
\begin{equation}
	\left( x + \lambda \right)^\alpha - x^\alpha \approx 
	\alpha \frac{\lambda}{x^{1 - \alpha}} = 2\pi; 
	\quad x = \frac{r}{r_0}.
\label{3-120}
\end{equation}
We suppose that the place where the SU(3) classical color field becomes quantum one is defined as the place where the quantum fluctuations in the volume $\Delta V = 4\pi r^2 \Delta r$ with 
\begin{equation}
	\frac{\Delta r}{r_0} \approx \lambda 
	\approx \frac{1}{\alpha} \frac{2 \pi}{x^{\alpha - 1}}
\label{3-130}
\end{equation}
of the corresponding field becomes comparable with magnitude of these fields 
\begin{equation}
	\Delta v \approx v, \quad 
	\Delta w \approx w
\label{3-140}
\end{equation}
Substituting of Eq's \eqref{3-40}, \eqref{2-60}-\eqref{2-80}, \eqref{3-100}, \eqref{3-110} , \eqref{3-130} and \eqref{3-140} into Eq.~\eqref{3-10} we obtain 
\begin{equation}
	\left( \frac{g'}{A} \right)^2 \approx 2 \pi
\label{3-150}
\end{equation}
where $\frac{1}{{g'}^2} = \frac{4 \pi}{g^2 \hbar c}$ is the dimensionless coupling constant that is similar to the fine structure constant in quantum electrodynamics $\alpha = \frac{e^2}{\hbar c}$. In quantum chromodynamics 
$\beta = 1/{g'}^2 \geq 1$. If we choose $1/{g'} \approx 1$ and from Fig.~\ref{v_r} we take $A \approx 0.4$ we see that 
\begin{equation}
	\left( \frac{g'}{A} \right)^2 \approx 6.25 
\label{3-160}
\end{equation}
that is comparable with $2 \pi \approx 6.28$. 
\par 
Thus in this section we have shown that if the condition \eqref{3-150} is true then at some distance from the center the transition from the classical phase to quantum one occurs. Unfortunately the rough estimation presented in this section does not allow us to calculate the radius where such transition takes place. For the exact evaluation of the place where such transition happens it is necessary to have \emph{non-perturbative} quantization methods which are missing at the moment. 

\section{The rotation curve of Yang-Mills colored dark matter}

In this section we would like to show that the solution of Eq's \eqref{2-40} \eqref{2-50} really has such mass density distribution that it is in a good agreement with the Universal Curve Rotation. 

\subsection{Numerical investigation}
\label{numerical}

\begin{figure}[h]
\begin{minipage}[t]{.45\linewidth}
  \begin{center}
  \fbox{
  \includegraphics[height=5cm,width=7cm]{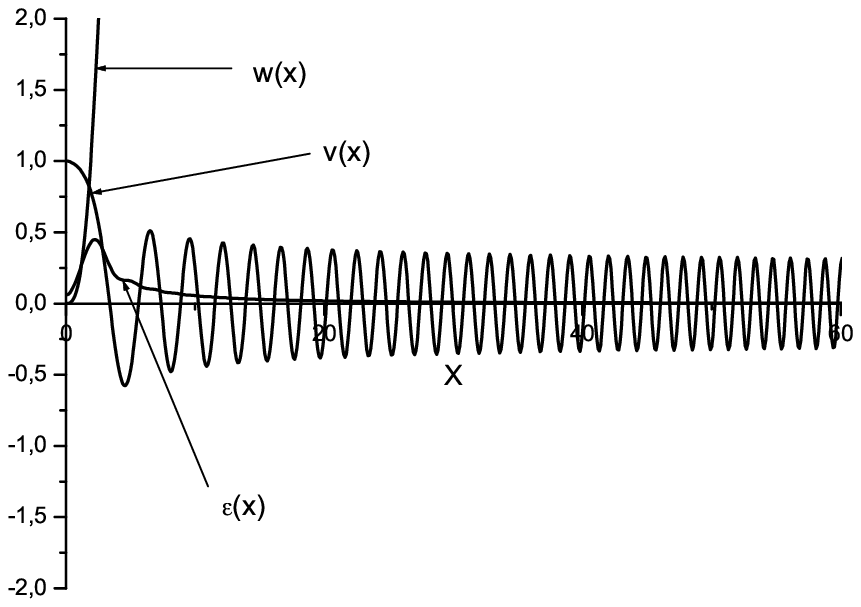}}
  \caption{The profile of functions $w(x), v(x), \varepsilon(x)$, $v_2=-0.1$, $w_3=1$.}
  \label{fg1}
  \end{center}
\end{minipage}\hfill
\begin{minipage}[t]{.45\linewidth}
  \begin{center}
  \fbox{
  \includegraphics[height=5cm,width=7cm]{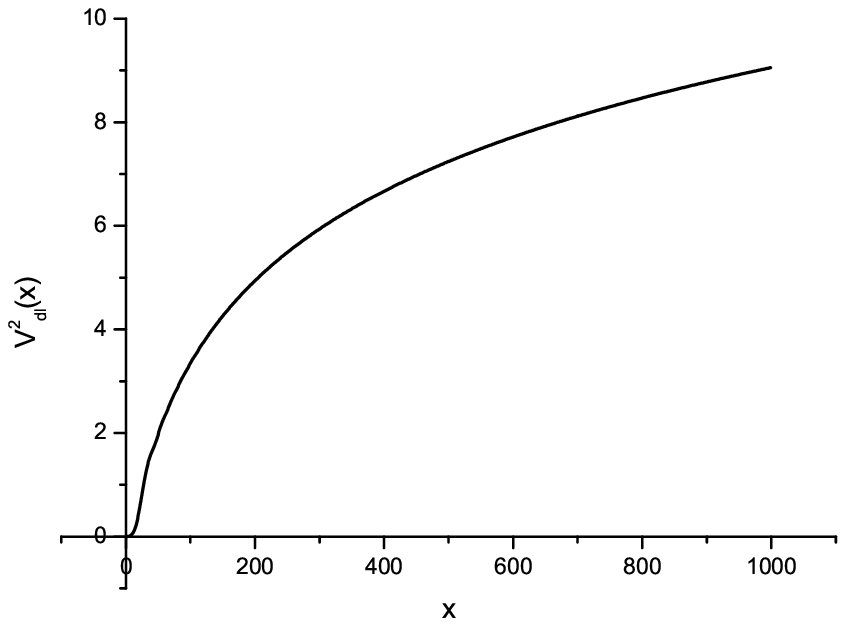}}
  \caption{The profile of the	dimensionless rotation curve 
  $V^2_{dl} = {g'}^2 \left( \frac{r_0}{l_{Pl}} \right)^2 \frac{V^2(x)}{c^2}$.}
  \label{fg2}
  \end{center}
\end{minipage}\hfill 
\end{figure}

In this subsection we present the typical numerical solution of Eq's \eqref{2-40} \eqref{2-50}. 
For the numerical investigation we have to start from the point $x = \delta \ll 1$. Here we have approximate solution \eqref{2-160} \eqref{2-170} and now we choose the parameter $r_0$ as 
\begin{equation}
	r_0 = \frac{1}{w_3^{1/3}}.
\label{5a-10}
\end{equation}
The typical behavior of functions $v(x)$ and $w(x)$ is presented in Fig.~\ref{fg1}. 
\par
The mass density $\rho(r)$ is 
\begin{equation}
	\rho(r) = \frac{1}{g^2 c^2 r_0^4} \rho(x)
\label{5a-20}
\end{equation}
where $\rho(x) = \varepsilon(x)$ and $\varepsilon(x)$ is given in Eq.~\eqref{2-90}. The profile of the dimensionless energy density  $\varepsilon(x)$ in Fig.~\ref{fg2} is presented.
\par
The rotation curve is defined as 
\begin{equation}
	V^2 = G \frac{\mathcal{M}(r)}{r} = 
	4 \pi G \frac{1}{r}
	\int \limits^r_0 r^2 \rho(r) dr = 
	\frac{G \hbar}{c^3} \frac{c^2}{{g'}^2 r_0^2} \frac{\mathcal{M}(x)}{x} = 
	\left[
		\frac{1}{{g'}^2} \left( \frac{l_{Pl}}{r_0} \right)^2 
		\frac{\mathcal{M}(x)}{x}
	\right] c^2 
\label{5a-30}
\end{equation}
where $\mathcal{M}(x)$ is the dimensionless mass of the color fields $A^a_\mu$ inside the sphere of radius $r = x r_0$, ${g'}^2 = g^2 c \hbar/ 4\pi$ is the dimensionless coupling constant, $G$ is the Newton gravitational constant. The parameter $\alpha \approx 1.31$ can be found using fitting of functions $w(x)$ or $\varepsilon(x)$, for details see Appendix~\ref{alpha}.

\subsection{The comparison with a Universal Rotation Curve of spiral galaxies}

\begin{figure}[h]
\begin{minipage}[t]{.45\linewidth}
  \begin{center}
  \fbox{
  \includegraphics[height=5cm,width=7cm]{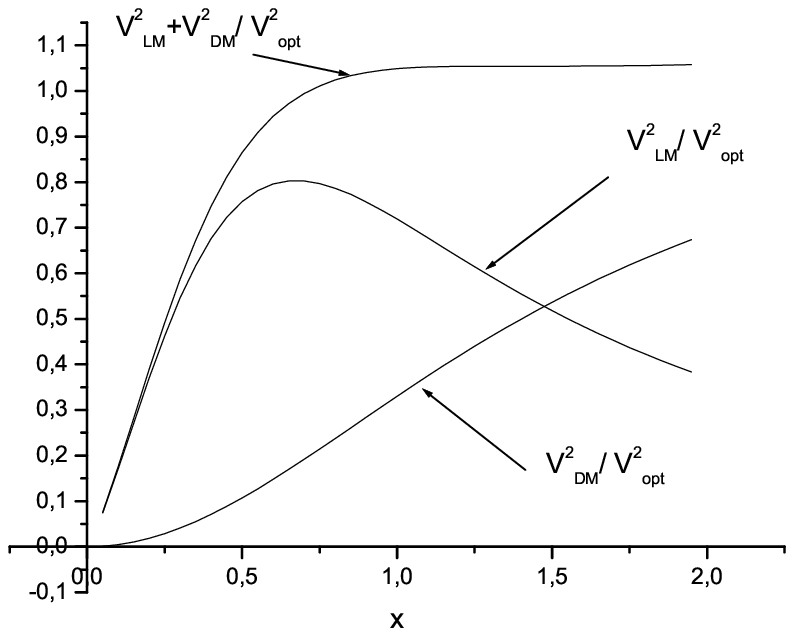}}
  \caption{The profiles of dimensionless rotation curves for the light and dark matter \cite{Persic:1995ru}.}
  \label{fg3}
  \end{center}
\end{minipage}\hfill
\begin{minipage}[t]{.45\linewidth}
  \begin{center}
  \fbox{
  \includegraphics[height=5cm,width=7cm]{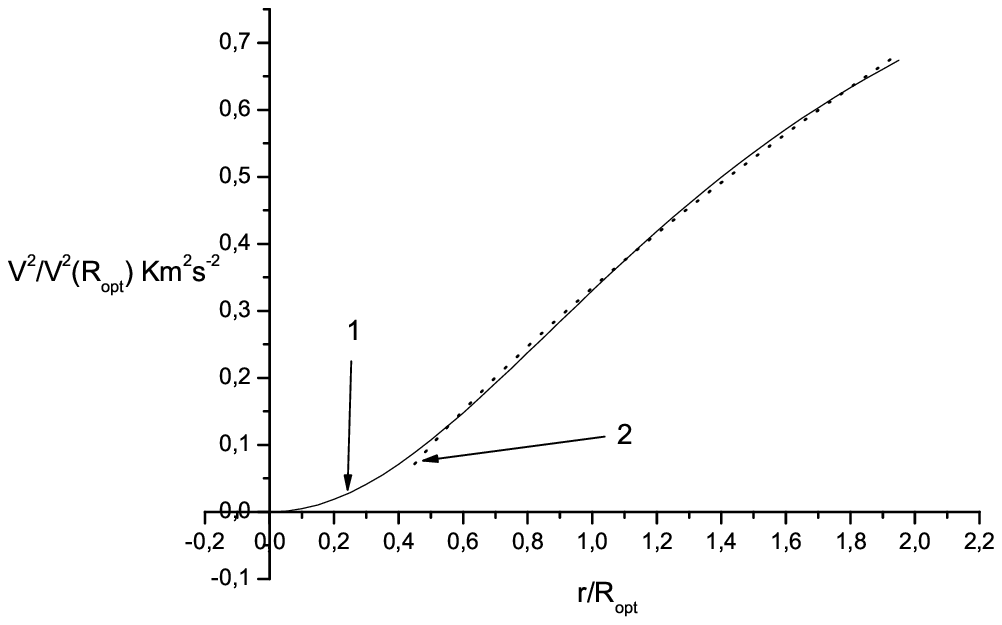}}
	  \caption{The comparison of DM rotation curve \eqref{5b-30} (curve~1) with the rotation curve \eqref{3-78} (curve~2) for the SU(3) classical color field \eqref{a1-10}-\eqref{a1-40}. 
	  $\alpha \approx 1.31, g' = 1, R_{opt} = 20 KPs = 6*10^{17} m, V_{opt} = 100 Km/s$.}
  \label{fg4}
  \end{center}
\end{minipage}\hfill 
\end{figure}

In Ref.~\cite{Persic:1995ru} a Universal Rotation Curve of spiral galaxies is offered that describes any rotation curve at any radius with a very small cosmic variance
\begin{equation}
	V_{URC} \left( \frac{r}{R_{opt}} \right) = 
	V(R_{opt}) \left[ 
		\left( 0.72 + 0.44 \log \frac{L}{L_*} \right) 
		\frac{1.97 X^{1.22}}{ \left( X^2 + 0.78^2 \right)^{1.43}} + 
		1.6\, e^{-0.4(L/L_*)} \frac{X^2}{X^2 + 1.5^2 
		\left( \frac{L}{L_*} \right)^{0.4}} 
	\right]^{1/2} {\rm km~s^{-1}}
\label{5b-10}
\end{equation}
where $R_{opt} \equiv 3.2\,R_D$ is the optical radius and $R_D$ is the disc exponential length-scale; $X = r/R_{opt}$; $L$ is the luminosity. We would like to compare the rotation curve for the color fields \eqref{5a-30} with the Universal Rotation Curve \eqref{5b-10} where, for example, $L/L_* = 1$
\begin{equation}
	V_{URC} \left( \frac{r}{R_{opt}} \right) = 
	V(R_{opt}) \left[ 
		\frac{1.4184 \; X^{1.22}}{ \left( X^2 + 0.78^2 \right)^{1.43}} + 
		\frac{1.07251 \; X^2}{X^2 + 1.5^2} 
	\right]^{1/2} {\rm km~s^{-1}}.
\label{5b-20}
\end{equation}
For the DM the Universal Rotation Curve is 
\begin{equation}
	V_{DM}^2 \left( \frac{r}{R_{opt}} \right) = 
	V^2(R_{opt}) \frac{1.07251 \; X^2}{X^2 + 1.5^2} \; 
	{\text{km}^2 \text{s}^{-2}}.
\label{5b-30}
\end{equation}
The profiles of $V_{URC}(X), V_{DM}^2(X), V_{LM}^2(X)$ in Fig.~\ref{fg3} are presented ($V_{DM}^2$ is the rotation curve for the DM, $V_{LM}^2(X)$ is the rotation curve for the light matter).
\par
At the center $r \approx 0$ the approximate solution has the form \eqref{2-160} \eqref{2-170} and the mass density \eqref{5a-20} approximately is 
\begin{equation}
	\rho(x) \approx \frac{2}{g^2 c^2 r_0^4} \left( 
		3 {v'_2}^2 + \frac{5}{54} x^2 
	\right).
\label{5b-40}
\end{equation}
Consequently the rotation curve will be 
\begin{equation}
	V^2(r) \approx c^2 \frac{1}{{g'}^2} \left( \frac{l_{Pl}}{r_0} \right)^2 
	\left[
		{v_2'}^2 \left( \frac{r}{r_0} \right)^2 + \frac{1}{54} {w'_3}^2 \left( \frac{r}{r_0} \right)^4 
	\right] \; \text{cm}^2 \text{s}^{-2}. 
\label{5b-50}
\end{equation}
The comparison with Eq.~\eqref{5b-30} by $x \ll 1$ gives us 
\begin{equation}
	\left( \frac{r_0}{l_{Pl}} \right)^4 \approx 10^{-6} 
	\left( \frac{c^2}{V^2} \right)^2 \frac{{v'}^2}{{g'}^2} 
	\left( \frac{ R_{Opt} }{l_{Pl}} \right)^2.
\label{5b-60}
\end{equation}
Far away from the center the dimensionless energy density $\varepsilon_\infty(x)$ is presented in Eq.~\eqref{2-100} and in this case we can estimate the values of square of speed in the following way 
\begin{eqnarray}
	V^2 &=& c^2 \frac{1}{{g'}^2} \left( \frac{l_{Pl}}{r_0^2}^2 \right) \frac{1}{x} 
	\left( 
		\int \limits_0^{x} x^2 \left[ \varepsilon(x) - \varepsilon_\infty(x) \right] dx + 
		\int \limits_{x_1}^x x^2 \varepsilon_\infty(x) dx
	\right) 
\nonumber \\
	&\approx& \left[
		c^2 \frac{1}{{g'}^2} \left( \frac{l_{Pl}}{r_0^2}^2 \right) \frac{1}{x} 
		\int \limits_0^x x^2 \varepsilon_\infty(x) dx 
	\right] - V^2_0 ,
\label{3-73}\\
	V^2_0 &=& c^2 \frac{1}{{g'}^2} \left( \frac{l_{Pl}}{r_0^2}^2 \right) \frac{1}{x}  \int \limits_0^{x} x^2 
	\left[ \varepsilon_\infty(x) - \varepsilon(x) \right] dx .
\label{3-76}
\end{eqnarray}
The numerical value of $V^2_0$ is defined near to the center of galaxy where according to Eq. \eqref{3-76} the difference $\varepsilon_\infty(x) - \varepsilon(x)$ is maximal. Thus the asymptotical behavior of the rotation curve for the domain filled with the SU(3) gauge field is 
\begin{equation}
	V^2 \approx \left[ 
	\frac{2}{3} \frac{1}{{g'}^2} \left( \frac{l_{Pl}}{r_0} \right)^2 
		\frac{\alpha^2 \left( \alpha - 1 \right) \left( 3 \alpha - 1 \right) }{2 \alpha - 1}
		\left( \frac{r}{r_0} \right)^{2 \alpha - 2} 
	\right]c^2 - V^2_0 .
\label{3-78}
\end{equation}
In Fig.~\ref{fg4} the profiles of the Universal Rotational Curve \eqref{5b-30} and fitting curve \eqref{3-78} are presented. The value of parameter $\alpha$ is given from section~\ref{numerical}. The details of fitting 
\begin{eqnarray}
	r_0 &\approx & 2.01 \cdot 10^{-18} \; \text{cm}, 
\label{3-81}\\
	V_0 &\approx & 32.25 \text{ Km} \cdot \text{s}^{-1}
\label{3-91}
\end{eqnarray}
in Appendix~\ref{asymptotic} are presented. One can see that the biggest disagreement is near to the center since close to the center the fitting curve have to be \eqref{5b-50} not \eqref{3-78}.

\subsection{Transition to non-perturbative quantized phase}

The energy density \eqref{2-90} gives us an infinite total mass. How can we avoid this problem ? In Section \ref{transition} we brought forward arguments that the gauge field $A^B_\mu$ inside of some region is in classical phase and outside the region is in quantum phase. We think that it is the manifestation of the fact that the gauge field is the strongly interacting field and the quantization of this field should be carried out using a non-perturbative technique. In fact in this paper we would like to show that non-perturbative quantized fields can be spatially distributed in such a way that classical and quantum phases exist simultaneously. 
\par
Unfortunately up to now we do not have any exact non-perturbative technique for the quantization. In this section we want to describe briefly approximate non-perturbative technique based on the Heisenberg approach \cite{heisenberg} for non-perturbative quantization of a nonlinear spinor field (for details, see \cite{Dzhunushaliev:2006di}). 
\par
In section \ref{transition} we have shown that at some distance from the center the classical phase changes on non-perturbative quantum phase. In Ref. \cite{Dzhunushaliev:2006di} it is shown that two scalar fields may describe a non-perturbative quantized gauge field. Briefly it can be shown by the following way. In quantizing strongly interacting SU(3) gauge fields - via Heisenberg's non-perturbative method \cite{heisenberg} one first replaces the classical fields by field operators 
$\mathcal A^B_{\mu} \rightarrow \widehat{\mathcal A}^B_\mu$. This yields the
following differential equations for the operators
\begin{equation}
    \partial_\nu \widehat {\mathcal F}^{B\mu\nu} = 0.
\label{sec2-31}
\end{equation}
where $\mu , \nu = 0,1,2,3$; $B=1,2, \cdots , 8$ are SU(3) color indices. These nonlinear equations for the field operators of the nonlinear quantum fields can be used to determine expectation values for the field operators
$\widehat {\mathcal A}^B_\mu$. One problem in using these equations in order to obtain expectation values like 
$\langle \mathcal A^B_\mu \rangle$, is that these equations involve not only powers or derivatives of $\langle \mathcal A^B_\mu \rangle$ ({\it i.e.} terms like 
$\partial_\alpha \langle \mathcal A^B_\mu \rangle$ or $\partial_\alpha
\partial_\beta \langle \mathcal A^B_\mu \rangle$) 
but also contain terms like 
$\mathcal{G}^{BC}_{\mu\nu} = \langle \mathcal A^B_\mu \mathcal A^C_\nu \rangle$. 
Starting with Eq. \eqref{sec2-31} one can generate an operator differential equation for the product $\widehat {\mathcal A}^B_\mu \widehat {\mathcal A}^C_\nu$ consequently allowing the determination of the Green's function $\mathcal{G}^{BC}_{\mu\nu}$
\begin{equation}
  \left\langle Q \left|
  \widehat {\mathcal A}^B(x) \partial_{y\nu} \widehat {\mathcal F}^{B\mu\nu}(x)
  \right| Q \right\rangle = 0.
\label{sec2-32}
\end{equation}
However this equation will in it's turn contain other, higher order Green's functions. Repeating these steps leads to an infinite set of equations connecting Green's functions of ever increasing order. This construction, leading to an infinite set of coupled, differential equations, does not have an exact, analytical solution and so must be handled using some approximation. The basic approach in this case is to give some physically reasonable scheme for cutting off the infinite set of equations for the Green's functions. Using some assumptions and approximations on 2- and 4-points Green's functions one can reduce the initial SU(3) Lagrangian to an effective Lagrangian describing two interacting scalar fields (for details see Ref. \cite{Dzhunushaliev:2006di}). The scalar fields $\phi$ and $\chi$ which are under discussion here appear in the following way. We assume that in the first approximation two points Green's functions can be calculated as follows 
\begin{eqnarray}
  \left\langle A^a_i (x) A^b_j (y) \right\rangle &=& - \eta_{i j} 
  f^{apm} f^{bpn} \chi^m(x) \chi^n(y),
\label{sec2-33}\\
	\left\langle A^a_0 (x) A^b_0 (y) \right\rangle & \ll & 
	\left\langle A^a_i (x) A^b_j (y) \right\rangle
\label{sec2-34}
\end{eqnarray}
where $A^a_\mu \in SU(2) \subset SU(3), a=1,2,3$; $m = 4,5,6,7,8$; $i,j = 1,2,3$ are spatial indices. And 
\begin{eqnarray}
  \left\langle A^m_i (x) A^n_j (y) \right\rangle &=& - \eta_{i j} 
  f^{mpa} f^{npb} \phi^a(x) \phi^b(y),
\label{sec2-35}\\
	\left\langle A^m_0 (x) A^n_0 (y) \right\rangle & \ll & 
	\left\langle A^m_i (x) A^n_j (y) \right\rangle
\label{sec2-36}
\end{eqnarray}
where $A^m_\mu \in SU(3)/SU(2)$. The 4-points Green's functions are a bilinear combination of 2-points Green's functions 
\begin{eqnarray}
  &&\left\langle A^m_\mu(x) A^n_\nu(y) A^p_\alpha(z) A^q_\beta(u) \right\rangle =
  \lambda_1 \biggl[ \left\langle A^m_\mu(x) A^n_\nu(y) \right\rangle 
	  \left\langle A^p_\alpha(z) A^q_\beta(u) \right\rangle + 
  \biggl.,
\nonumber \\
	  &&
 	 \biggl.
	  \frac{\mu_1^2}{4} \left(
	  	\delta^{mn} \eta_{\mu \nu} \left\langle A^p_\alpha(z) A^q_\beta(u) \right\rangle + 
	  	\delta^{pq} \eta_{\alpha \beta} \left\langle A^m_\mu(x) A^n_\nu(y) \right\rangle
	  \right) + 
	  \frac{\mu_1^4}{16} 
	  \delta^{mn} \eta_{\mu \nu} \delta^{pq} \eta_{\alpha \beta} 
	  \biggl] + 
\nonumber \\
	  &&
	  ( \text{permutations of indices} )
\label{sec2-37}
\end{eqnarray}
and 
\begin{eqnarray}
  &&\left\langle A^a_\mu(x) A^b_\nu(y) A^c_\alpha(z) A^d_\beta(u) \right\rangle =
  \lambda_2 \biggl[ \left\langle A^a_\mu(x) A^b_\nu(y) \right\rangle 
	  \left\langle A^c_\alpha(z) A^d_\beta(u) \right\rangle + 
  \biggl.
\nonumber \\
	  &&
	  \biggl.
	  \frac{\mu_2^2}{4} \left(
	  	\delta^{ab} \eta_{\mu \nu} \left\langle A^c_\alpha(z) A^d_\beta(u) \right\rangle + 
	  	\delta^{cd} \eta_{\alpha \beta} \left\langle A^a_\mu(x) A^b_\nu(y) \right\rangle
	  \right) + 
	  \frac{\mu_2^4}{16} \delta^{ab} \eta_{\mu \nu} \delta^{cd} \eta_{\alpha \beta}
	  \biggl] + 
\nonumber \\
	  &&
	 \text{(permutations of indices)}
\label{2f1-30}
\end{eqnarray}
here $\lambda_{1,2}, \mu_{1,2}$ are some constants. The assumptions \eqref{sec2-33}-\eqref{2f1-30} allows us to average the SU(3) Lagrangian 
\begin{equation}
  \mathcal{L}_{SU(3)} = - \frac{1}{4}	F^A_{\mu \nu} F^{A \mu \nu}, \;
  A = 1,2, \cdots , 8
\label{2f-40}
\end{equation}
and bring it to the form 
\begin{eqnarray}
  \mathcal{L}_{eff} &=& \left\langle  \mathcal{L}_{SU(3)} \right\rangle = 
  \frac{1}{2} \left( \partial_\mu \phi^a \right) 
    \left( \partial^\mu \phi^a \right) 
    + \frac{1}{2}  \left( \partial_\mu \chi^m \right) 
    \left( \partial^\mu \chi^m \right) - V(\phi^a, \chi^m) ,
\label{2f-50}\\
	V(\phi^a, \chi^m) &=& \frac{\lambda_1}{4} \left(
        \phi^a \phi^a - \mu_1^2 
    \right)^2 - 
    \frac{\lambda_2}{4} \left(
        \chi^m \chi^m - \mu_2^2 
    \right)^2 - \frac{1}{2} \left( \phi^a \phi^a \right) \left( \chi^m \chi^m \right) 
\label{2f-55}
\end{eqnarray}
with the field equations 
\begin{eqnarray}
	\nabla_\mu \left( 
		\nabla^\mu \phi^a
	\right) &=& - \frac{\partial V\left( \phi^a, \chi^m \right)}{\partial \phi^a} ,
\label{sec2f-60}\\
	\nabla_\mu \left( 
		\nabla^\mu \chi^m 
	\right) &=& - \frac{\partial V\left( \phi^a, \chi^m \right)}{\partial \chi^m} .
\label{sec2f-70}
\end{eqnarray}
Let us consider the spherically symmetric case $\phi^a = k \phi(r), \chi^m = k \chi(r)$ where $k$ is some constant. In this case the field equations are 
\begin{eqnarray}
	\frac{d^2 \phi}{d r^2} + \frac{2}{r} \frac{d \phi}{d r} &=& \phi \left[
		\chi^2 + \lambda_1 \left( \phi^2 - \mu_1^2 \right)
	\right] ,
\label{sec2f-80}\\
	\frac{d^2 \chi}{d r^2} + \frac{2}{r} \frac{d \chi}{d r} &=& \chi \left[
		\phi^2 + \lambda_2 \left( \chi^2 - \mu_2^2 \right)
	\right] .
\label{sec2f-90}
\end{eqnarray}
It is easy to see that asymptotically the solution has the form 
\begin{eqnarray}
	\phi(x) &\approx& m_1 + 
	\phi_\infty \frac{e^{-\left( r - r_q \right) \sqrt{2 \lambda_1 \mu_1^2}}}{r}  ,
\label{sec3-30}\\
	\chi(x) &\approx& \chi_\infty \frac{e^{-\left( r - r_q \right) \sqrt{m_1^2 - \lambda_2 \mu_2^2}}}{r} 
\label{sec3-40}
\end{eqnarray}
where $\phi_\infty, \chi_\infty, r_q$ are constants. We think that this solution describes the non-perturbative quantized SU(3) gauge field after the transition from classical phase to quantum one occurs. 
\par 
The main point of this consideration is that the non-perturbative quantized gauge field decreases very quickly (exponentially) after transition to the quantum phase and consequently the total mass becomes finite one. 

\section{Invisibility of colored dark matter}

For the detection of DM (in the context of the model presented here) we can use only \textit{colored} particles which can interact with color gauge fields. The equations describing the motion of a colored particle are Wong's equations 
\begin{eqnarray}
	m \frac{d^2 x^\mu}{ds^2} &=& -g F^{A \mu}_\nu Q^A \frac{d x^\nu}{ds} ,
\label{4-10}\\
	\frac{d Q^A}{ds}  &=& -g f^{ABC} \left(
		A^B_\mu \frac{d x^\mu}{ds} 
	\right) Q^C
\label{4-20}
\end{eqnarray}
where $x^\mu(s)$ is the 4D trajectory of the particle with the mass $m$, $Q^A$ is the color  components of color charge of the particle, $\left( Q^A \right)^2 = Q^2 = const$. The ordinary elementary particles are colorless and consequently do not interact with the color DM. 
\par
Only 't-Hooft~-~Polyakov monopoles and dyons may interact with colored DM and can be used for the detection of the colored DM. Another possibility of the interaction of an elementary particle with colored DM is the interaction between external color field (DM) and a color electric and/or magnetic dipole or quadrupole of this elementary particle. 

\section{Conclusions}

In this paper we have considered the idea that the problem of DM probably can be connected with the problem of non-perturbative quantization of strongly interacting fields. It allows us to connect one problem in macroscopical physics (the problem of DM) with another problem in microscopical physics (confinement problem on quantum chromodynamics). In this connection R. Kolb in Ref.~\cite{Kolb} write: ``Dark matter and dark energy are two of the
binding cords I will use to illustrate how collaborations of astronomers and high energy physicists on large astronomical projects can be good for astronomy, and how discoveries in astronomy can guide high-energy physicists in their quest for understanding nature on the smallest scales.''. 
\par 
The features of the model of colored DM presented here are: 
\begin{itemize}
	\item The estimation of the parameter $\alpha$ from the Yang-Mills equations (see Appendix \ref{alpha}) and from the astrophysical (see Appendix \ref{asymptotic}) point of view is in agreement with remarkable accuracy. 
	\item On the background of a non-perturbative vacuum of non-Abelian gauge field there exists a bubble of the same  classical gauge field. The classical non-Abelian gauge field filled the bubble is colored DM.
	\item Spherically symmetric classical solutions of the Yang-Mills equations have weakly decreasing mass density distribution leading to a good agreement with the Universal Rotation Curve. 
	\item The distribution of these classical gauge field is that on some distance from the center the transition from classical phase to quantum occurs.
\end{itemize}
The problems for the future investigations are:
\begin{itemize}
	\item The theoretical estimation of $V_0^2$ parameter and comparing it with the fitted value \eqref{3-91}. 
	\item The fitting of the Universal Rotational Curve \eqref{5b-30} using the function joining \eqref{5b-50} and \eqref{3-78}. 
	\item The most important problem in the model presented here is the calculation of gauge field distribution using a non-perturbative quantization technique. 
	\item The search for possibility of the classical gauge fields detection  . 
\end{itemize}

\section*{Acknowledgements}

I am very grateful for P. Kozlov for the help of fitting. 
\appendix 

\section {Ansatz for SU(3) gauge potential in Minkowski spacetime} 

We consider the classical SU(3) Yang-Mills gauge field $A^B_\mu$ and use the following ansatz for the $SU(2) \in SU(3)$ components of the gauge field \cite{corrigan} 
\begin{eqnarray}
	A_0^2 &=& - 2 \frac{z}{gr^2} \chi(r), \quad
	A_0^5 = 2 \frac{y}{gr^2} \chi(r), \quad
	A_0^7 = - 2 \frac{x}{gr^2} \chi(r), 
\label{a1-10}\\
	A^2_i &=& 2 \frac{\epsilon_{3ij} x^j}{gr^2} \left[ h(r) + 1 \right] ,
\label{a1-20}\\
	A^5_i &=& -2 \frac{\epsilon_{2ij} x^j}{gr^2} \left[ h(r) + 1 \right] ,
\label{a1-30}\\
	A^7_i &=& 2 \frac{\epsilon_{1ij} x^j}{gr^2} \left[ h(r) + 1 \right] 
\label{a1-40}
\end{eqnarray}
where $F^B_{\mu \nu} = \partial_\mu A^B_\nu - \partial_\nu A^B_\mu + g f^{ABC} A^B_\mu A^C_\nu$ is the field strength tensor; $f^{ABC}$ are the SU(3) structural constants; $A,B,C = 1, 2, \cdots , 8$ are color indices; $g$ is the coupling constant and 
\begin{eqnarray}
	\left( A_0 \right)_{\alpha , \beta} &=& 2 \left( 
		\frac{x^\alpha x^\beta}{r^2} - \frac{1}{3} \delta^{\alpha \beta}
	\right) \frac{w(r)}{gr} ,
\label{a1-50}\\
	\left( A_i \right)_{\alpha \beta} &=& 2 \left(
		\epsilon_{is \alpha} x^\beta + \epsilon_{is \beta} x^\alpha
	\right) \frac{x^s}{gr^3} v(r) ,
\label{a1-60}
\end{eqnarray}
for the coset components belonging to the coset space $SU(3)/SU(2)$; $i=1,2,3$ are space indices; $\epsilon_{ijk}$ is the absolutely antisymmetric Levi-Civita tensor; the functions $\chi(r), h(r), w(r), v(r)$ are unknown functions. The coset components $\left( A_\mu \right)_{\alpha \beta}$ in the matrix form are written as 
\begin{equation}
	\left( A_\mu \right)_{\alpha \beta} = 
	\sum \limits_{a=1,3,4,6,8} A_\mu^B \left( T^B \right)_{\alpha , \beta} 
\label{a1-70}
\end{equation}
where $T^B = \frac{\lambda^B}{2}$ are the SU(3) generators, $\lambda^B$ are the Gell-Mann matrices. 
\par 
This ansatz in the spherical coordinate system is 
\begin{eqnarray}
	A^a_t &=& \left\{
		w(r) \sin^2 \theta \sin(2 \varphi);   \quad 
		-2 \phi(r) \cos \theta ;   \quad 
		w(r) \sin^2 \theta \cos (2 \varphi);   \quad
		w(r) \sin(2 \theta) \cos \phi ;  
	\right.
\nonumber \\
	&&
	\left.
		2 \phi(r) \sin \theta \sin \varphi;  \quad
		w(r) \sin(2 \theta) \sin \phi;  \quad
		- \phi(r) \sin(2 \theta);  \quad
		w(r) \frac{1 + 3 \cos(2 \theta)}{2 \sqrt{3}}
	\right\};
\label{a1-80}\\	
	A^a_r &=& 0;
\label{a1-90}\\
	A^a_\theta &=& \left\{
	-2  v(r)\cos \left(2\varphi\right) \, \sin \theta ; \quad
	0; \quad
	2  v(r)\sin \theta \sin\left( 2\varphi \right) ;  \quad
	2  v(r)\cos \theta \sin \varphi ;  \quad
	2 \left[ 1 + h(r) \right] \cos \varphi ;  \quad
	\right.
\nonumber \\
	&&
	\left.
	-2 v(r) \cos \theta \cos \varphi ;  \quad
	2 \left[ 1 + h(r) \right] \sin \varphi ;  \quad
	0 
	\right\};
\label{a1-100}\\
	A^a_\varphi &=& \left\{
		v(r) \sin(2 \theta) \sin(2 \varphi) ;   \quad
		-2 [1 + h(r)] \sin \theta ;   \quad
		v(r) \sin(2 \theta) \cos (2 \varphi ) ;  \quad
		2 v(r) \cos(2 \theta) \cos \varphi ;   \quad
	\right.
\nonumber \\
	&&
	\left. 
	-2 [1 + h(r)] \cos \theta \sin \varphi ;   \quad
	2 v(r) \cos (2 \theta) \sin \varphi ;   \quad
	2 [1 + h(r)] \cos \theta \cos \varphi ;   \quad
	\sqrt{3} v(r) \sin (2 \theta) 
	\right\} .
\label{a1-110}
\end{eqnarray}

\section{Fitting of parameter $\alpha$}
\label{alpha}

For the estimation of the parameter $\alpha$ we use the functions $w(x)$ and $\varepsilon(x)$ given from the numerical solution of Eq's~\eqref{2-40}~\eqref{2-50} (see also Fig.~\ref{fg1}) in the region $20 \leq x \leq 100$. The fitting functions are the asymptotical form of the function $w(x)$ (see Eq.~\eqref{2-70}) 
\begin{equation}
	w(x) \approx \alpha x^\alpha + w_0 
\label{a2-10}
\end{equation}
and for the control we use the asymptotical form of the function $\varepsilon(x)$ (see Eq.~\eqref{2-100}) 
\begin{equation}
	\varepsilon_\infty(x) \approx \frac{2}{3} \alpha^2 \left( \alpha - 1 \right) 
	\left( 3 \alpha - 1 \right) x^{2\alpha - 4} + \varepsilon_0 
\label{a2-20}
\end{equation}
here $w_0$ and $\varepsilon_0$ are systematical errors in the consequence of ignoring of other components in the asymptotical decompositions \eqref{a2-10} and \eqref{a2-20}. 
\par
The fitting carried out using MATHEMATICA package. The fitting parameters are: $\alpha$ and either $w_0$ or $\varepsilon_0$. The result of fitting is 
\begin{equation}
	\alpha \approx 1.31077 , \quad 
	w_0 \approx 2.15369 
\label{a2-30}
\end{equation}
for fitting $w(x)$ and 
\begin{equation}
	\alpha \approx 1.31995 , \quad 
	\varepsilon_0 \approx -0.0012265 
\label{a2-40}
\end{equation}
for fitting $\varepsilon(x)$. 

\section{Fitting of rotational curve of gauge field}
\label{asymptotic}

For the fitting of the rotational curve \eqref{3-78} we use the data from the Universal Rotational Curve \eqref{5b-30}. The fitting equation is equation \eqref{3-78} in the form 
\begin{eqnarray}
	\frac{V^2}{V^2_{opt}} &=& A x^B + C, 
\label{a3-02}\\
	A &=& \frac{2}{3} 
	\frac{\alpha^2 \left( \alpha - 1 \right) \left( 3 \alpha - 1 \right) }{2 \alpha - 1}
	\frac{c^2}{V^2_{opt}}
	\frac{1}{{g'}^2} \left( \frac{l_{Pl}}{r_0} \right)^2 
	\left( \frac{R_{opt}}{r_0} \right)^{2 \alpha - 2} 
\label{a3-04}\\
	B &=& 2 \alpha - 2, 
\label{a3-06}\\
	C &=& - V_0^2
\label{a3-08}
\end{eqnarray}
where the fitted parameters are $A,B,C$. The fitting is carried out using MATHEMATICA package. The fitting parameters are: $r_0$ and $V^2_0$. The result of fitting is 
\begin{equation}
	\alpha = 1.31954, \quad 
	r_0 \approx 2.01 \cdot 10^{-18} \; \text{cm}, \quad 
	V_0 \approx 32.25 \text{ Km} \cdot \text{s}^{-1}. 
\label{a3-10}
\end{equation}


\begin{thebibliography}{99}

\bibitem{yang_mills}
C. N. Yang and R. L. Mills, 
Phys. Rev., \textbf{96}, 191 (1954). 

\bibitem{higgs}
P. W. Higgs, 
Phys. Lett., \textbf{12}, 132 (1964). 

\bibitem{thooft}
G. 't Hooft, 
Nucl. Phys., \textbf{B35}, 167 (1971). 

\bibitem{thooft2}
G. 't Hooft, 
Nucl. Phys., \textbf{B79}, 276 (1974). 

\bibitem{polyakov}
A. M. Ployakov, 
Phys. Lett., \textbf{B59}, 82 (1975). 

\bibitem{bpst}
A. A. Belavin, A. M. Polyakov, A. S. Schwarz and Yu. S. Tyapkin, 
Phys. Lett., \textbf{B59}, 85 (1975). 

\bibitem{Obukhov:1996ry}
Y.~N.~Obukhov: 
``Analogue of black string in the Yang-Mills gauge theory,'' 
\textit{Int.\ J.\ Theor.\ Phys.},  Vol. 37, (1998), 1455 - 1468; 

\bibitem{Dzhunushaliev:1999fy}
V.~D.~Dzhunushaliev and D.~Singleton: 
``Confining solutions of SU(3) Yang-Mills theory,'' In: Dvoeglazov, V.V. (Ed.): 
\textit{Contribution to Contemporary Fundamental Physics}, Nova Science Publishers, 1999, pp. 336-346.

\bibitem{corrigan}
E.~Corrigan, D.~I.~Olive, D.~B.~Farlie and J.~Nuyts: 
``Magnetic monopoles in SU(3) gauge theories'', 
\textit{Nucl.~ Phys.,} Vol. B106, (1976), pp.475-492.

\bibitem{Dzhunushaliev:2006di}
V.~Dzhunushaliev, ``Color defects in a gauge condensate,''
\textit{Preprint}:hep-ph/0605070.

\bibitem{Persic:1995ru}
M.~Persic, P.~Salucci and F.~Stel,
``The Universal rotation curve of spiral galaxies: 1. The Dark matter connection,''
\textit{Mon.\ Not.\ Roy.\ Astron.\ Soc.,}  Vol. 281, (1996) pp.27 - 47,

\bibitem{heisenberg}
W. Heisenberg, 
\textit{Introduction to the unified field theory of elementary particles.},
(Max - Planck - Institut f\"ur Physik und Astrophysik, Interscience
Publishers London, New York-Sydney, 1966).

\bibitem{Kolb}
Rocky Kolb, ``A Thousand Invisible Cords Binding Astronomy and High-Energy Physics'', 
astro-ph/0708.1199.

\end{thebibliography}
\end{document}